\documentclass[fleqn,12pt,twoside]{article}
\usepackage{wrapfig}
\usepackage{wrapft}
\usepackage[headings]{espcrc1}

\readRCS
$Id: espcrc1.tex,v 1.2 2004/02/24 11:22:11 spepping Exp $
\usepackage{graphicx}
\newcommand{\be}{\begin{eqnarray}}
\newcommand{\ee}{\end{eqnarray}}

\newcommand{\AmS}{{\protect\the\textfont2
  A\kern-.1667em\lower.5ex\hbox{M}\kern-.125emS}}

\title{Toward the theory of strongly coupled Quark-Gluon Plasma
 (sQGP) }

\author{Edward Shuryak
\address{Department of Physics and  Astronomy,\\
University at Stony Brook, NY 11794, USA}
}

\runtitle{sQGP}
\runauthor{E.Shuryak}

\begin{document}

% typeset front matter
\maketitle

\begin{abstract}
We review recent progress toward understanding of sQGP. The
phenomenological part includes discussion of elliptic and conical flows
at RHIC. Then we proceed to first quantum mechanical studies of manybody
states at $T>Tc$, the ``polymeric
chains'' $\bar q.g.g... q$ and baryons. A new model for sQGP is
a classical dynamical system, in which  color vector is changed
via the Wong equation. First Molecular Dynamics (MD) results
for its diffusion and viscosity are reported.
Finally we speculate how strong correlations in matter may help
solve puzzles related to jet quenching, both the magnitude and
angular distribution.  
\end{abstract}

\section{Introduction}
\begin{figure}
\includegraphics[width=8cm]{uinfty.eps}
\includegraphics[width=8cm]{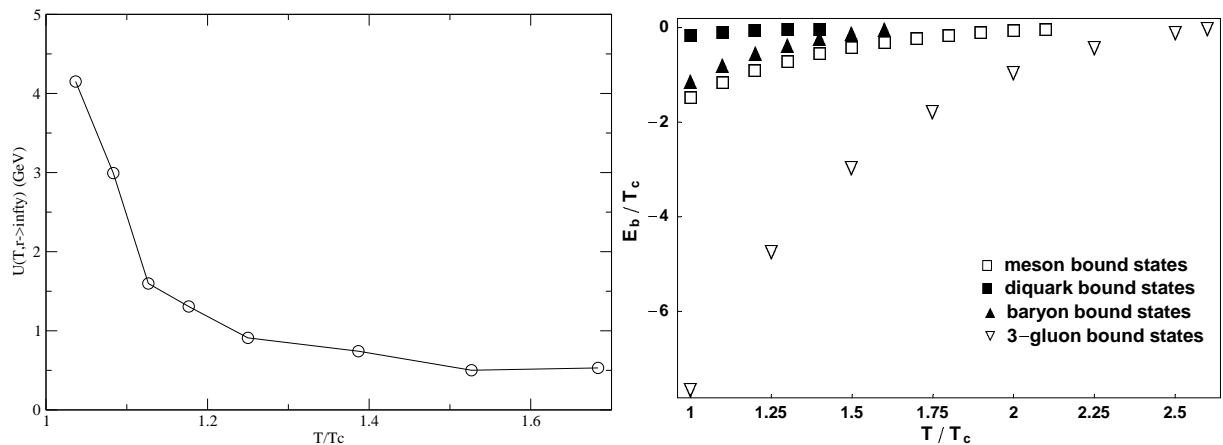}
 \caption{\label{fig_boundstates}
(a)
The separation energy $U(T,r\rightarrow\infty)-U(T,r=.2 fm)$ in $GeV$ vs $T/T_c$,
calculated from the free energy  \cite{potentials} by removing the entropy term.(b)
 Dependence of various states' binding energy (in units of
$T_c$) on the temperature.
 }
 \end{figure}

The idea that QGP at RHIC, $T=(1-2)T_c$, seems to be in a
strongly coupled  regime (sQGP) was introduced in 2003
\cite{Shu_liquid,SZ_rethinking,SZ_CFT}  and was discussed a lot at
the ``discovery''  BNL workshop \cite{discovery_workshop} and QM04.
Let me start with
 the simplest physics point possible.  Although at
 $T>T_c$ quarks are deconfined, the energy needed for separation
of quarks close to
 $T_c$ is huge (see
Fig.\ref{fig_boundstates}a, up to $U\sim 4\, GeV$. 
%. Although at $T>?? T_c$ it is getting small and
%negative,
%as Debye theory would say $U(r=\infty,T)=const g^3 T, const<0$, it is
%positive and truly
%huge near $T_c$,  $U(r=\infty,T_c)/2\approx 2\, GeV$ per charge.
One implication is simple:
 the charges simply cannot get separated until very
 high $T$. The second:
% if e.g. this energy is due to formation of a
% pair of B-type mesons, we have matter-induced effective mass of a quark
%up to about 1 GeV. Last but not least:
% anyone who tries to understand the  sQGP structure
%  has to
% deal with
as the  ratio $U/T\sim 10$
 goes into the Boltzmann exponents, any perturbative
approach is completely hopeless.

Instead of going to many other arguments, let me simply list main
 reasons  which demand such
a radical change, from traditional weak coupling to
 strong coupling methods:\\
\noindent
1.Collective phenomena observed at RHIC lead the hydro practitioners to view
 QGP as a ``near perfect liquid'' \cite{Shu_liquid,Teaney_visc};  \\
2.Feshbach-type resonances  due to marginal states may lead to
 large cross sections \cite{SZ_rethinking}.\\
3.Classical
  e/m plasma  can be a good liquid too, if sufficiently strongly coupled.\\
4. A close relative of QCD, the 
 $\cal N$=4 supersymmetric Yang-Mills gauge theory can be studied
in a strongly coupled $g^2N_c\rightarrow \infty$ regime at finite $T$ via
the AdS/CFT correspondence: the results are very close to what we
 observe at RHIC.

Since 
I cannot cover all these topics in depth, let me just provide
a more detailed list of these arguments, with at
 least some statements and references:\\
\noindent
{\bf(1a)} The departure
of elliptic flow data from the hydro prediction happens only at rather high
$p_t\sim 1.5-2\, GeV$, from which the estimated 
{\em viscosity-to-entropy ratio}
$\eta/s=.1-.2$ \cite{Teaney_visc} is more than order of magnitude lower
than in pQCD.  \\
{\bf(1b)} Another transport coefficient, the charm diffusion constant $D_c$
deduced from single electron $R_{AA}
$ and $v_2$ (much discussed at this meeting),
is also an order of magnitude lower than pQCD estimates
\cite{MT}.\\
{\bf(1c)} New hydrodynamical phenomenon suggested recently \cite{CST},
the {\em conical} flow, maybe explains why secondaries from a
  quenched jet fly preferentially to large angle
 $\approx 70$ degrees
 consistent with the Mach angle for (time-averaged) speed of sound.\\
%
%\noindent
{\bf(2a)} Combining lattice data on quasiparticle masses and
interparticle
potentials, one indeed finds a lot of bound states  \cite{SZ_bound}
and  resonances \cite{Mannarelli:2005pz} at $T>Tc$\\
{\bf(2b)} The same approach explains why $\eta_c,J/\psi$ remains bound
till near $3T_c$, as was directly observed on the lattice \cite{charmonium};\\
{\bf(2c)} this approach was experimentally demonstrated to work for
ultra-cold trapped fermionic atoms $Li^6$, turning it to near prefect
liquid as well when the Feshbach resonance leads to the scattering length
$a\rightarrow \infty$.  Experiments  on oscillations found a
sharp minimum in damping near the resonance, reducing it by about 2 orders
of magnitude. According to our study \cite{GSZ} two lowest modes
are well described by hydro with {\em ``quantum viscosity''} $\eta\approx
0.3\hbar n $, in a way nearly as low as that of sQGP.\\
{\bf(2d)}  Heavy-light
resonances in sQGP
can  explain the value of the charm diffusion constant \cite{HGR}.\\
%
%\noindent
{\bf(3a)} The interaction parameter  
$\Gamma=<potential\,\, energy>/T$
in sQGP is obviously not small, O(10).
At such  $\Gamma$ the  classical strongly coupled
  e/m plasma is a good liquid: we find the same in
our classical version of sQGP as well;\\
%
%\noindent
{\bf(4a)} The EoS for finite $T$ $\cal{N}$=4 SUSY YM 
is similar to what is seen on the lattice in the RHIC domain, namely
$p/p_{ideal}=[(3/4)+O((g^2N_c)^{-3/2})]$ \cite{thermo};\\ 
{\bf(4b)}At infinite coupling
there is a finite limit of viscosity  $\eta/s=>1/4\pi$ \cite{PSS}, again
close
to the RHIC value. The  ``R-charge'' diffusion constant is low as well.\\
{\bf(4c)} One may even think about AdS/CFT ``gravity dual'' to the
whole
RHIC collisions, with black hole production \cite{Nastase} which 
then fly away from the
test brane \cite{SSZ}.

\section{Systematics of collective flows}
\subsection{Elliptic flows}
Collective flows observed at SPS and RHIC are quite accurately reproduced
by ideal hydrodynamics for the QGP/mixed phase, complemented by
a hadronic cascade for the
hadronic phase \cite{ourhydro}.  The radial flow predictions are
 in quantitative agreement with the data
for all secondaries, from $\pi$ to $\Omega^-$. 
 Elliptic flow is described by $v_2=<cos(2\phi)>$ depending on 6 variables 
 $v_2(s,pt,M_i,y,b,A)$,
the energy, transverse momentum, particle mass, rapidity, centrality
and system size. Hydro works 
for not-too-large $p_t<2 GeV$  (above it a different
and still poorly understood regime starts.)\\
(i) The energy dependence was the issue I discussed already at QM99\cite{ES99},
 $v_2(s)$ was predicted to be smoothly rising from SPS to RHIC by
about factor 2, as stiffer EoS of the QGP replaces
that of hadronic gas and mixed phase, see more in \cite{ourhydro}. It 
 was soon confirmed by the first RHIC data\footnote{
Hydro with fixed freezeout, at $T_f$, lead to a non-monotonous curve
which is not the case.}.
 \\
(ii) the (pseudo)rapidity dependence $v_2(\eta)$ 
has a triangular shape (PHOBOS). 
Although the first 3-d hydro by Hirano et al were not
able to reproduce it,  now with
 a hydro+cascade approach  \cite{Hirano_now} it works well. 
The physics is basically the same as 
for $v_2(s)$: in fact $v_2$ displays good ``limiting
fragmentation'', it depends mostly on local density 
$v_2\left(dN/dy(s,y)\right)$ 
rather than on $s$ and $y$ by itself.\\
(iii)
 The $v_2(p_t,M_i)$ dependence
for different  species, known as the ``fine structure'' of the
elliptic flow, was discussed here
by R.Lacey \cite{Lacey_now} who pointed out a scaling
relation $ v_2\sim M_i y_t^2 $ where $y_t$ is the transverse
rapidity.  Suggested by hydro, it is perfectly  satisfied by the data
for all secondaries.\\
(iv)  Centrality and size dependence: ideal hydro
is scale invariant and 
$v_2$ is basically given by the spatial deformation, thus
 $v_2(b,A)/\epsilon_2(b,A)\approx const(b,A)$. This relation is in
excellent agreement with the data.

  Remaining hydro-skeptics  include
 Bhalerao et al \cite{B3O}, who
ascribe the rise in $v_2(s)$ (or that in  $v_2(\eta)$ toward
 midrapidity) to
an ``incomplete equilibration''. It was pointed out earlier
by H.Heselberg et al \cite{uli_etal} 
that hydro scaling  (iv) should be
violated for sufficiently peripheral collisions, by a  ``dilute regime''. 
Unfortunately,
  due to
  experimental difficulties  that regime was never really
clearly observed. Bhalerao et al pointed out possible dependence on 
the system size, 
predicting  violation of hydro scaling
$v_2(CuCu)/v_2(AuAu)\sim 1/3$. Recent  $CuCu$ data do not
  agree with them while the hydro scaling is still satisfied. 

 An overall blue sky still has two clouds:
(i) $v_4/v_2^2$ seem to be a constant, but not .5 as ideal hydro 
predicts, but about 1.2. It may be related to the issue of
$v_2$
fluctuations discussed by Mrowczynski and myself \cite{Mrowczynski:2002bw}.).
(ii) the HBT radii remain a puzzle.  Paradoxically.
we can better describe early
stages  (when the elliptic flow is formed) than
very late dilute ones.
It may well be that our freezeout conditions via
naive cascades are no good:
effective potentials, resonances etc  may be
different than we thought\footnote{One recent attempt, by Cramer and
 G.Miller,
has generated  discussion here. I complained that they do not
account for the  pions absorbed by
their optical potential on the way out. 
B.Muller, in a summary, defended them
because ``the source term in the HBT
formalism
describes the vertex position of the last inelastic interaction''.
As the author of that formalism (in 1973) I cannot agree more:
 thus  Cramer and
  Miller better follow their pions to that $last$ interaction.
 }.  

\subsection{Conical flow}

 Let me start  reminding a general
 kinematics of the most
elementary process $1=>2$, possible  for a jet in matter.
 No change in the mass implies
$0=(p-k)^2-p^2=-2(pk)+k^2$.  Provided the last small term
(recoil) can be dropped,
 the  angle between $\vec k$ and
$\vec p$ is then
 $cos(\theta)=(k_0 p_0/k p) $, For very fast particle $p_0/ p=1 $
and thus the condition for the process is simply $k_0<k$.

  Now, the RHIC data on 2-particle correlation with a triggered jet
show that most radiation goes into a peak at angles 60-70 degrees,
so that $cos(\theta)\approx 1/3$ and thus that must be the $k_0/k$
ratio for what is emitted. This fits very well to sound waves,
for which this ratio is the
speed of sound $c_s $, and its average value over the duration
of the collision is indeed about\footnote{Note, we don't mean here
  speed of sound squared! }  1/3. One simple way to test that was suggested
by Antinori and myself \cite{Antinori:2005tu}: if the jet is a b quark
(which can be tagged experimentally) $p_0/p=1/v>1$ which causes the
cone
to shrink, till it goes to zero at the critical velocity
$v=c_s$. Gluon radiation behaves in the opposite way with $decreasing$
$v$, and never shrinks to zero.

%%%%%%%%%%%%%%%%%%%%
\begin{wrapfigure}{l}{6.cm}
\begin{minipage}{6.cm}
\centering
   \begin{center}
      \includegraphics[height=5.cm,width=5.cm]{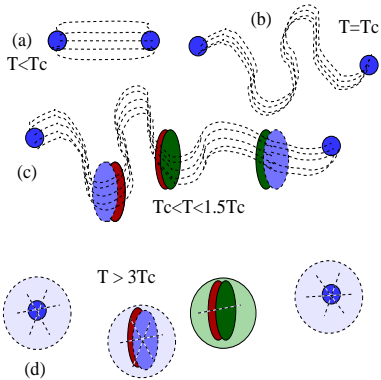}
   \end{center}
   \caption{
   \label{fig_fields}
  Schematic picture of matter, as the temperature grows
   }
\end{minipage}
\end{wrapfigure}
%%%%%%%%%%%%%%%%%%%%

 A number of people suggested  a possibility of gluon
 Cherenkov radiation, see the talk by Ruppert for details.
There is 
 very little support
in data or theory for that: but even the optimists find
 the angles for Cherenkov emission
of gluons with $p\sim 1 \, GeV$ to be at least order of magnitude
smaller than the observed
peak at $\sim 70$ degrees.

  Whether  the mechanism of $dE/dx$ is  a gluon radiation or
  scattering
or ionization, a small mean free path implies rapid local deposition of energy,
with  subsequent hydrodynamical process.
  This issue is discussed in detail in the talk of my collaborator
 Casalderrey-Solana
  \cite{CST}: The observed shape of the peak, its magnitude and $p_t$
dependence is in rough agreement with the data. The issue of
flow subtraction seem to be resolved (see B.Cole's talk), but 
 3-particle
correlations are still too controversial to comment.
Let me only  remind that our
 hydro solutions  are obtained with
arbitrary viscosity, and that the observed effects already
  constrain the viscosity  more than all  data on $v_2$ since
 the gradients  are larger  for
 conical than for elliptic flow.

\section{First quantum manybody states above $T_c$}
 The existence of bound states of quark and gluon quasiparticles
at $T>T_c$, suggested in \cite{SZ_rethinking}
 was much discussed at this meeting. 
 Let me first report on recent studies
 \cite{LS}, done in collaboration with
grad.student
J.F.Liao.

One result is about {\em polymeric chains} of the type
 $\bar q g g ..g q$ which appear naturally in a string
 picture\footnote{Recall that a gluon has two color indices and can
 thus be connected to two strings.}.
 For this problem 
the most useful coordinates in this case are not the usual Jacobi
but ``chain coordinates'', using which we 
have  proven that such
polymers have the same binding energy {\em per bond}\footnote{Not per
particle, so  long enough chains have twice more binding per particle
 than mesons.}
as for $\bar q q$ mesons.
We also studied (variationally) 
two 3-body problems: the  closed (3-)chains of gluons
$(ggg)$
and  $(qqq)$  baryons. Their binding energy (and range) is shown in 
Fig.\ref{fig_boundstates}: one can see that the former one is quite
robust
while baryons have quite marginal binding (because of smaller relative
charge
of quarks).
Before we go forward with a general discussion, let us try to
summarize
the proposed scenario as a single picture, see Fig.\ref{fig_boundstates}(b).
From relatively short string-like configuration of color fields at low
$T$,
fig (a), one moves to longer strings (b) at the critical point
\cite{Polyakov:1978vu}. New is picture (c) which depicts ``polymeric
chains'' considered in this work, significant at $T=(1-1.5) T_c$.  
Eventually, at high $T$, one goes into (d) with independent quark and
gluon
quasiparticles, neutralized by isotropic Debye clouds.

%%%%%%%%%%%%%%%%%%%%
\begin{wrapfigure}{l}{6.cm}
\begin{minipage}{6.cm}
\centering
   \begin{center}
      \includegraphics[height=11.cm,width=6.cm]{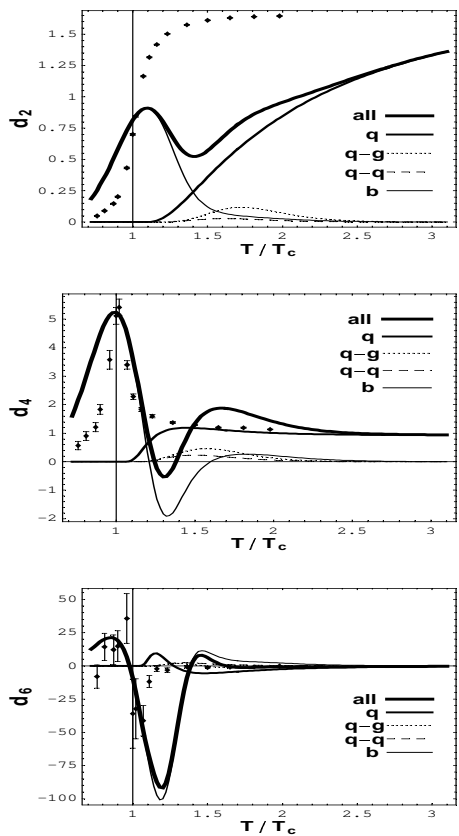}
   \end{center}
 \caption{\label{fig_c4_baryons}
The contribution of different states to (a) $d_2$, (b)  $d_4$ and (c)
$d_6$
in a ``bound state'' model, the points are data
\protect\cite{ci_paper}}
\end{minipage}
\end{wrapfigure}
%%%%%%%%%%%%%%%%%%%%

Let me now briefly touch on criticism of the ``bound state QGP''
raised at this meeting. First, let us be clear what is being
criticized. The bound states are not  {\em ad hoc} ideas 
 but the direct
 mathematical  consequence of: (i)
 quasiparticle masses  from
 the lattice \cite{masses}; (ii)  lattice-based effective interaction
between colored charges \cite{potentials}; and (iii) quantum mechanics.
 None of the
  critics ever suggested  that either
 (i) or (ii) or  (iii) may be wrong. Where the direct test 
has been made, for the charmonium states \cite{charmonium}, one finds a
good agreement. 
And it does not take  much  to realize that
quasiparticles with $M\sim 1\, GeV$ should behave like
the charmed quarks.

  V.Koch et al \cite{Baryon_Strangeness} and
 F.Karsch\cite{Karsch_talk} argued that they do not see 
a significant contributions of diquarks (and baryons) in
the lattice data on  $ d_n(T)= {\partial^n (p/T^4)/ \partial
  (\mu/T)^n}|_{\mu=0}$ with n=2,4,6 at $T>1.2 Tc$.
It is hardly surprising, as there are not so many states
of those and the expected masses are rather large,
$2M_q$ and   $3M_q$, respectively
\footnote{In fact all their arguments can be repeated verbatim
for say $T\sim 50 MeV$ or so, with the same conclusion: 
no visible baryons in a hadronic gas. They do exist as states there,
of course.}
. 
%To put the discussion below
% into proper perspective, keep in mind that  
%there should be 3 categories of the expected bound states, 
%in decreaseing robustness, 
%: (i) glueballs, (ii)
%$(qg)_3$
%and mesons $\bar q q$; and (iii) $(qg)_6$, diquark and
% baryons. Denoting
% the strength
%of the effective potential in $\bar q q$ states counted as 1, these categories
%have the strength of interaction 9/4-2, 9/8-1 and 1/2, respectively.
%Thus the baryon number carrying states we will discuss in this work
%belong to the $third$ category, the
% most  weakly bound once and thus naturally suspect.
(In \cite{SZ_bound} we have not even included diquarks and baryons in pressure,
to which they clearly contribute too little.) 
 Karsch \cite{Karsch_talk} further  argued  that the ratio
$d_4/d_2\approx <B^2>$ gives directly  the mean constituent baryon number $B$,
and  since this ratio gets close to 1 above $T_c$, all 
 bound states with $B>1$ are excluded.
But the same reasoning gives
 $d_6/d_4\approx <B^2>$ as well: this ratio is however
not close to 1 but $\sim -10$.
 Does it exclude any quark gas model 
as well? 

In fact the argument is simply too naive and ignores a lot of things.
What is worse, Karsch et al have not explained the most prominent
features of his data, the
peak in $d_4(T)$ and the ``wiggle' in  $d_6(T)$. see  Fig.\ref{fig_c4_baryons}, or large
flavor-nondiagonal $d_4(T),d_6(T)$.

 It was  pointed out in ref.\cite{BKS}, that since 
the  quasiparticle masses may depend on $\mu$ their
 derivatives such as $M''=(\partial^2 M/\partial \mu^2)|_{\mu=0}$ must be 
included in all formulae. The same is true for masses of the bound
states such as baryons.
It is shown in our paper  \cite{LS_2} that all above mentioned features
of higher susceptibilities
are naturally  explained by the expected change in
the baryon ($N,\Delta$) mass 
 $M_B(T,\mu)$. At the phase boundary it is expected to grow
from the ``vacuum mass'' $\sim 1 GeV$ to much larger
value in sQGP, $\approx 3M_q$. As a result  $M_B''$
is large and  changes sign at the
 inflection point near $T_c$: 
 that is why there is  a peak in $d_4(T)$ and the ``wiggle' in  $d_6(T)$,
 see the baryon curve $b$ in  Fig.\ref{fig_c4_baryons}.
(The effect of binding of $qq$ and $qg$ states is not included in
this plot: in the latter case it can increase its contribution by
factor 2-3 and make a better agreement with data, especially for $d_2$.)

\section{Strongly coupled
  colored classical plasma, studied via  Molecular Dynamics}
 Nice introduction to strongly coupled Abelian e/m plasmas has been
provided here by M.Thoma, so
let me jump directly to the point.
 The interaction parameter is defined as
\be \Gamma = C_c \alpha_s n^{1/3} N_{corr}/T\ee
where the color Casimir is  $C_c=4/3$ and $3$ for
$q$ and $g$.  The quasiparticle
density\footnote{
Quasiparticles inside various bound states should also be included.} 
is proportional to large number of degrees of freedom in QGP $n^{1/3}\sim
N_{dof}^{1/3} T$ where  $N_{dof}=16+2*2*3*N_f\sim 50$. 
 The number of ``strongly correlated  partners'' $N_{corr}$ is just 1
for binary states, 2 for polymeric chains we discussed above, but goes up to
12 for a fcc crystal (see below). Combining all, even for $\alpha_s\sim 1$ 
one finds $\Gamma>1$, possibly up to $O(10)$. Further, the
structure is exponentially sensitive to $\Gamma$ as it goes to
Boltzmann factor. And, as we show below,
$\Gamma\sim 10 $ is precisely where the plasma liquid is most perfect.

 New model B.Gelman, I.Zahed and myself
are proposing \cite{GSZ_cQGP} 
  is based on the
following main assumptions:
(i) The particles are heavy enough to move non-relativistically,
with masses\footnote{Recall that close to $T_c$ the $M/T$ ratio for
quasiparticles reaches 4-5. } $M>>T$; 
(ii) Their interaction is dominated by 
colored electric (Coulomb) interactions,
with all magnetic effects (spin forces, etc) ignored; 
(iii) Their color representations are large, so that
color operators $t^a$ can be represented by 
 classical vectors.

Dynamics of color vectors as well as particle coordinates 
are described by classical
equations of motion (EoM)\footnote{Another
model, amenable to MD simulations, has been proposed at this meeting
by the Budapest group \cite{Budapest}: but the color part is not classical
and is subject to complicated stochastic transitions. Not only color is
not
conserved in their model, but (as far as i can tell), the energy cannot be
 conserved either.
} known as Wong equation\footnote{It can also be rewritten as 
a set of canonically conjugated equations, for one x-p pair for SU(2)
and
3 pairs for SU(3), see \cite{CManuel} for refs.
 B.Muller in summary 
expressed doubts whether this model is gauge invariant:
 no reason to worry: it obviously is, this is how
Wong  derived it.
}.
\be m{dC^a\over d\tau}= g f^{abc} p_\mu A_\mu^b C^c\ee
where $A_\mu$ is a local field induced by other particles.
 The total color and total energy
are thus conserved (which is monitored during actual MD calculations). 
The interaction potential
is proportional to the dot product of the unit color vectors
%indicating color directions 
%\be A_0^{\alpha,i,\beta,j}=( Q_{\alpha\,i}^a\,Q_{\beta\,j}^a)V(R_{ij}) \ee
%Here $R_{ij}=|x_{\alpha\,i}-x_{\beta\,j}|$ is the interparticle
%distance, $i,j$ and $\alpha,\beta $ are the particle indices and the
%types ($\bar q, q, g$).
times the potential. The latter
eventually should be  deduced from lattice simulations, we so far use
just a Coulomb.
In order to stabilize the system, we have also added a short-range
repulsion which is suppose to mimic
 the quantum-mechanical localization energy
$V_{short}=\hbar^2 /M r^2$, like it is sometimes done to
describe systems like solid $He$.

The main variable parameter of the model is obviously 
the temperature $T$. Eventually, in applications to sQGP all 4
parameters  mentioned above are in fact some functions of $T$,
so such applications of the model in the narrow sense 
 would only be restricted to some 1-d line in a 4-d parameter
space. But in order to understand them better
 it would be necessary first to describe  the properties of the
model in a wider sense, 
in all its parameter space.

  The closest physical e/m analog of our model is the classical 
two-component ionic plasma, e.g. a molten  ionic salts\footnote{It is a good approximation to think
 an electron to be completely transfered
from $Na$ to $Cl$ and $T$ still low enough not
to excite other electrons.} 
such as $NaCl$. 
For large  $\Gamma\rightarrow\infty$ this system freezes into the
 fcc cubic lattice
with alternating positive and negative charges. 
For smaller 
$1<\Gamma<\Gamma_c\sim 80$ one deals with a strongly coupled ionic
liquid.
In the non-Abelian case one
also expects that the system gets frozen at very large $\Gamma$, with
the same {\em cubic crystal} plus quasi-Abelian
{\em ferromagnetic} (alternating) order
of the color vectors. 
%%%%%%%%%%%%%%%%%%%%
\begin{wrapfigure}{l}{6.cm}
\begin{minipage}{6.cm}
\centering
   \begin{center}
\vskip .3cm
\includegraphics[width=6.cm]{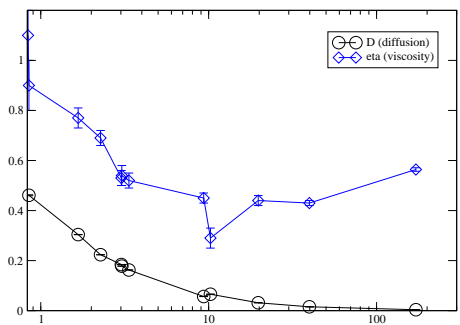}
   \end{center}
 \caption{\label{fig_eta_d}
%(a) A sketch of expected behavior of particle diffusion
%coefficient $D$ and viscosity $\eta$ on the interaction strength
%parameter $\Gamma$; (b)
 Our first MD data for the SU(2) classical plasma: 
 the particle diffusion
coefficient $D$ and viscosity $\eta$ 
versus $\tilde
\Gamma=<|Pot.energy|>/<Kin.energy>$.  
}
\end{minipage}
\end{wrapfigure}
%%%%%%%%%%%%%%%%%%%%

  Molecular dynamics (MD)
simulations can follow evolution of classical systems, weak or strong coupling.
Our studies of that are at exploratory stage as of now,
done so far with $4^3$ and $6^3$ particles with the SU(2) color vector 
(a vector on a 3-sphere).
 In the QM05 talk
I have shown some movies of melting  a crystal 
 by ``heating'' a system with a small random force:
a bit of friction can cool it back. Even when
pictures look homogeneous and the color vector looks isotropic,
there are local correlators
 of neighboring particles at large $\Gamma$.

Our interest is not so far in detailed studies of its microscopic
structure
per se, but in collective excitations -- the phonons and plasmons,
as well as their widths reflecting transport properties such as the
(self) diffusion coefficient $D$ and viscosity $\eta$ by Kubo 
formulae\footnote{
In fact MD is well suited for such studies, while in lattice QCD the
 Euclidean time makes it next to impossible 
to use Kubo-like formulae and access such transport properties.}. 
%A qualitative sketch of expected dependence on these quantities
%on the interaction
%parameter is shown in Fig.\ref{fig_eta_d}(a).
 With increased potential energy (or decreased $T$) the
particle mobility decreases, together with $D$. But viscosity has a
more interesting trend, with a minimum and subsequent rise toward a
``glassy'' liquid and solid regimes. This happens because
the momentum transfer can be achieved not only by particle propagation 
but also by phonons/plasmons, which have large mean free path
in perfect solids. Thus  an optimum exists, with
 ``the most perfect liquid'': 
 as  Fig.\ref{fig_eta_d} shows, it is at $\Gamma\sim 10$ as well.

\section{New views on jet quenching and charm trapping}
 The relevance of substructure of sQGP should be even more
important for transport properties.
 It was first suggested by  Zahed and myself
 in \cite{SZ_rethinking}
that existence of marginal states must increase rescattering and
thus dramatically reduce viscosity (mean free paths), leading to a
collisional (hydrodynamical) regime of expansion. Since different
states get marginal at different $T$, one may hope this mechanism
to work at all $T$ up to about $2T_c$, the highest temperature at
RHIC.

 We will now argue that the situation can be different for
 jet quenching. Its radiative theory  tells that all properties
of matter come one combination,  $\hat q = 5-15 \,GeV^2/fm$, the
values  currently under discussion. In most papers this value is
 expressed as the gluon density, with $dN/dy=1000-3000$, and some people 
argued this may be too large and contradict to entropy produced, see e.g.
\cite{Muller:2005en}. However  the relation between $\hat q$ and $dN/dy$
assumed is simply invalid in the sQGP regime.
(Think of a $NaCl$-like local structure, with alternating charges:
the electric field between the ions is a coherent sum of fields
from positive and negative charges, which increases local fields and
 $\hat q$
compared to randomly placed charges.)

  Another way to see why jet quenching (and charm equilibration) should be
enhanced near $T_c$  is to think about collective momentum/energy
sharing in  multibody correlated clusters we studied above. In particular,
an admixture of polymeric 
 chains  is known  in material science to be very effective mechanism
of the momentum
distribution over larger volume of matter.
A famous example is 
  Kevlar fibers added to epoxy (or other plastic),
 now applied in wide range
of applications, from tires, boats etc. to such exotic ones as
 ``bullet-proof  vests'' and even   ``anti-mine boots''.

 If this is so, there are consequences which can immediately be
 checked
in experiment.
If indeed only a ``well correlated'' (polymerized?) sQGP provides
large
dE/dx, it should only happen at $T$ rather close to  $T_c$.
Since at RHIC
the initial temperature is  about $2T_c$, jets may propagate with
 smaller losses,   till
the matter cools down and correlates properly.  Such a delay
 would  affect the  angular distribution
of the jet quenching phenomenon for non-central collisions.

In fact the original idea of ``jet
tomography'' via jet quenching is  in serious trouble
for  quite a while, because
the most natural assumption --
the
energy loss  proportional to the matter density --
is in strong contradiction with
 the observed strong angular dependence of jet quenching \cite{v2_largept},
predicting  too weak azimuthal asymmetry for non-central
collisions. It was however recently pointed out by Pantuev
\cite{Pantuev:2005jt} that a better description of data can be
achieved $if$ the jet quenching at the highest RHIC energy is
switched on after some ``latent time'' of about 2.2 fm. This time
quite reasonably matches a cooling time from $T\approx 2 T_c$ to
$T\approx 1.5 T_c$ at RHIC.

\section{Status of the theory}

  Studies of sQGP are
developing very rapidly, with  many new 
 ideas and connections to other fields  emerging daily.
 Instead of a summary,
let me comment instead on the rather complex status of QGP theory in general.

  Before RHIC
 there were many approaches which
looked reasonable, but then RHIC data put it under severe tests.
Most of them have  not survived, failing to
 explain  collective  flows, especially the elliptic $v_2$\footnote{ 
 String models (RQMD etc) get it factor 3-4 too small, and
 predicted its $decrease$ with $s$. ``Naive'' parton models like HIJING
 even predicted a bit negative values, and the classical color glass fields
got the $p_t$ dependence wrong. Parton cascade from Duke 
never even reported their $v_2$, as far as I know.}.  It took the courage
of Gyulassy and Molnar 
 \cite{GM} to tell us that a parton cascade can only get close to
the data if the cross sections 
are huge,  $\sigma_{gg}\sim 30 mb$. But if so,
the number of particles interacting at any moment
is large and  the very  concept of a cascade -- free motion
interrupted by scattering events -- looses its meaning\footnote{ Molecular
Dynamics takes its place in the
  classical context.}. Lattice studies have found
very large potentials between colored charges
near and above $T_c$. Now a challenge to all of us is to learn how
such a strongly coupled system of quasiparticles can be organized.
 The paradigm shift, from wQGP to sQGP, had definitely occurred.
Of course, all these events created a
lot of stress in theory circles. 

  Is the lesson  learned? Well,   
 a lot of people are in denial. 
%hoping that a miracle will 
%some version 
%of weak couplin methods, the  parton cascades  or bottom-up
%equilibration,
%can mimic 
%the hydro regime 
% or describe recent data on charm.
%
Example: in the theory summary  B.Muller, after briefly
 acknowledging
 a discovery of ``near perfect liquid'', proceeds to
 ``new ideas'' such as a parton cascade with plasma instabilities.
This subject, reviewed by  Mrowczynski,
is indeed interesting, and recently numerical studies
have shown how such instability may develop in a quasi-Abelian regime. But now
we are approaching its moment of truth: its time to do simulation 
without boxes, with appropriate initial 
distribution of partons, testing whether the instabilities can indeed 
lead to a ``little bang'' and $v_2$. I may be wrong, but
 there are many reasons to think that a collisionless
weakly coupled plasma will fail this test.
How large gluon masses and potentials seen on the lattice
can be put into negatives by feeble magnetic effects? How can those 
 explain heavy charm quark
 stopping/flow, much discussed at this meeting? 

  The situation is well
 illustrated  
 by a cartoon shown in the summary by B.Muller\footnote{Although
he used it in a somewhat
 different context, I think subconsciously he had my interpretation in
 mind.}, displaying
 a chicken afraid to cross the street. Yes, crossing from
familiar wQGP to the sQGP  is scary. 
My advice: don't be  a chicken, learn to fly, be an eagle.
 High above there is no fear, and one can see what people do
 in other fields. And, last but not least, there is basically no
 alternative:
one can sell non-working theories only for so long.


\begin{thebibliography}{99}
\itemsep -2pt 

\bibitem{Shu_liquid}
%\bibitem{Shuryak:2003xe}
  E.V.Shuryak,
  ``Why does the quark gluon plasma at RHIC behave as a nearly ideal fluid?,''
  Prog.\ Part.\ Nucl.\ Phys.\  {\bf 53}, 273 (2004)
  [ hep-ph/0312227].
  %%CITATION = HEP-PH 0312227;%%

\bibitem{SZ_rethinking}
E.V.Shuryak and I. Zahed, {\tt hep-ph/0307267},
Phys.\ Rev.\ C {\bf 70}, 021901 (2004)

\bibitem{SZ_CFT}
E.V.Shuryak and I. Zahed,
Phys.\ Rev.\  {\bf D69} (2004) 014011.
[ hep-th/0308073].
%%CITATION = HEP-TH 0308073;%%

\bibitem{discovery_workshop}
M.~Gyulassy and L.~McLerran,
  %``New forms of QCD matter discovered at RHIC,''
  Nucl.\ Phys.\ A {\bf 750}, 30 (2005)
  [ nucl-th/0405013].
  %%CITATION = NUCL-TH 0405013;%%
  E.~V.~Shuryak,Prog.Part.Nucl.Phys.53:273-303,2004, hep-ph/0312227
   %%CITATION = HEP-PH 0312227;%%
  %``What RHIC experiments and theory tell us about properties of quark-gluon
  %plasma?,''
  Nucl.\ Phys.\ A {\bf 750}, 64 (2005).
\bibitem{potentials}
O.~Kaczmarek, S.~Ejiri, F.~Karsch, E.~Laermann and F.~Zantow,
{\tt hep-lat/0312015}.

\bibitem{Teaney_visc} 
%\bibitem{Teaney:2003kp}
  D.~Teaney,
  %``The effect of viscosity on spectra, elliptic flow, and HBT radii,''
   nucl-th/0301099.
  %%CITATION = NUCL-TH 0301099;%%
 {\tt nucl-th/0301099}
\bibitem{MT}
  G.~D.~Moore and D.~Teaney,
 % ``How much do heavy quarks thermalize in a heavy ion collision?,''
   hep-ph/0412346.
  %%CITATION = HEP-PH 0412346;%%
 also D.Teaney's talk here.

\bibitem{CST}
%\bibitem{Casalderrey-Solana:2004qm}
  J.~Casalderrey-Solana, E.~V.~Shuryak and D.~Teaney,
  %``Conical flow induced by quenched QCD jets,''
   hep-ph/0411315.
  %%CITATION = HEP-PH 0411315;%%
see also Casalderrey-Solana's talk

\bibitem{SZ_bound}E.~V.~Shuryak and I.~Zahed,
%``Towards a theory of binary bound states in the quark gluon plasma,''
Phys.\ Rev.\ D {\bf 70}, 054507 (2004), hep-ph/0403127.
%%CITATION = HEP-PH 0403127;%%

\bibitem{Mannarelli:2005pz}
  M.~Mannarelli and R.~Rapp,
  %``Hadronic modes and quark properties in the quark-gluon plasma,''
   hep-ph/0505080.
  %%CITATION = HEP-PH 0505080;%%

\bibitem{charmonium}
S.~Datta, F.~Karsch, P.~Petreczky and I.~Wetzorke,
%``A study of charmonium systems across the deconfinement transition,''
{\tt hep-lat/0208012}. 
M. Asakawa and T. Hatsuda,
Nucl. Phys. {\bf A715} (2003) 863c; 
%``$J/\psi$ and
%$\eta_c$ in the deconfined plasma fom lattice QCD,"
hep-lat/0308034;

\bibitem{GSZ}
 B.~A.~Gelman, E.~V.~Shuryak and I.~Zahed,
  ``Cold Strongly Coupled Atoms Make a Near-perfect Liquid,''
   nucl-th/0410067.
  %%CITATION = NUCL-TH 0410067;%%

\bibitem{HGR}H.~van Hees, V.~Greco and R.~Rapp,
  %``Heavy-Quark Probes of the Quark-Gluon Plasma at RHIC,''
   nucl-th/0508055.
  %%CITATION = NUCL-TH 0508055;%%
\bibitem{thermo}
G.T. Horowitz and A. Strominger, { Nucl. Phys.} {\bf B360} (1991) 197.
S.S.Gubser, I.R.Klebanov and A.A. Tseytlin, {Nucl.\ Phys.\ } {\bf B534} (1998) 202
\bibitem{PSS}
G.~Policastro, D.~T.~Son and A.~O.~Starinets,
%``The shear viscosity of strongly coupled N = 4 supersymmetric Yang-Mills  plasma,''
Phys.\ Rev.\ Lett.\  {\bf 87} (2001) 081601.
%CITATION = HEP-TH 0104066;%%
\bibitem{Nastase}H.~Nastase,
  %``The RHIC fireball as a dual black hole,''
  hep-th/0501068.
  %%CITATION = HEP-TH 0501068;%%
\bibitem{SSZ} E.V.Shuryak,S.J.Sin,I.Zahed,in progress
\bibitem{ES99}
E.~V.~Shuryak,
  %``What have we learned and want to learn from heavy ion collisions at  CERN
  %SPS?,''
  Nucl.\ Phys.\ A {\bf 661}, 119 (1999)
  [ hep-ph/9906443].
  %%CITATION = HEP-PH 9906443;%%


\bibitem{ourhydro}
D. Teaney, J. Lauret, and E. V. Shuryak, Phys. Rev. Lett. {\bf
86}, 4783 (2001); 
\bibitem{Hirano_now} T.Hirano's talk at this meeting
\bibitem{Lacey_now} R.Lacey's talk at this meeting
\bibitem{B3O}
  R.~S.~Bhalerao, J.~P.~Blaizot, N.~Borghini and J.~Y.~Ollitrault,
  %``Elliptic flow and incomplete equilibration at RHIC,''
   nucl-th/0508009.
  %%CITATION = NUCL-TH 0508009;%%
\bibitem{uli_etal}
%P. F. Kolb, P. Huovinen, U. Heinz, and H.
Heiselberg, Phys. Lett. B {\bf 500}, 232 (2001).
     
\bibitem{Mrowczynski:2002bw}
  S.~Mrowczynski and E.~V.~Shuryak,
  %``Elliptic flow fluctuations,''
  Acta Phys.\ Polon.\ B {\bf 34}, 4241 (2003)
  [ nucl-th/0208052].
  %%CITATION = NUCL-TH 0208052;%%

\bibitem{Antinori:2005tu}
  F.~Antinori and E.~V.~Shuryak,
  %``A comment on conical flow induced by heavy-quark jets,''
   nucl-th/0507046.
  %%CITATION = NUCL-TH 0507046;%%

\bibitem{LS}
  J.~Liao and E.~V.~Shuryak,
  %``Polymer chains and baryons in a strongly coupled quark-gluon plasma,''
   hep-ph/0508035.
  %%CITATION = HEP-PH 0508035;%%
\bibitem{CManuel}P.F. Kelly, Q. Liu, C. Lucchesi, C. Manuel, Phys.Rev.D50:4209-4218,1994; hep-ph/9406285 

\bibitem{Polyakov:1978vu}
  A.~M.~Polyakov,
  %``Thermal Properties Of Gauge Fields And Quark Liberation,''
  Phys.\ Lett.\ B {\bf 72}, 477 (1978).
  %%CITATION = PHLTA,B72,477;%%
\bibitem{masses}
P. Petreczky, F. Karsch, E. Laermann,
S. Stickan, I. Wetzorke,  Nucl. Phys. Proc. Suppl. {\bf 106} (2002) 513.
\bibitem{Baryon_Strangeness}V. Koch, A. Majumder, and J. Randrup, {\tt nucl-th/0505052}.
\bibitem{Karsch_talk}  S.Ejiri,F.Karsch
and K.Redlich, hep-ph/0509051, F.Karsch, talk at this meeting.
\bibitem{ci_paper}
C.~R.~Allton {\it et al.},
  %``Thermodynamics of two flavor QCD to sixth order in quark chemical
  %potential,''
  Phys.\ Rev.\ D {\bf 71}, 054508 (2005)
  [arXiv:hep-lat/0501030].
  %%CITATION = HEP-LAT 0501030;%%
\bibitem{LS_2} J.~Liao and E.~V.~Shuryak, What do lattice baryonic
  susceptibilities  tell us about quarks,  diquarks and baryons at
  $T>Tc$?
 hep-ph/0510110
\bibitem{BKS}M.~Bluhm, B.~Kampfer and G.~Soff,
  %``Quasi-particle model of strongly interacting matter,''
  J.\ Phys.\ G {\bf 31}, S1151 (2005)
  [ hep-ph/0411319].
  %%CITATION = HEP-PH 0411319;%%

\bibitem{GSZ_cQGP}B.~A.~Gelman, E.~V.~Shuryak and I.~Zahed, in progress
\bibitem{Budapest} P.Hartmann, this proceedings.

\bibitem{v2_largept} E.V. Shuryak,
%THE AZIMUTHAL ASYMMETRY AT LARGE P(T) SEEM
%    TO BE TOO LARGE FOR A `JET QUENCHING'.
    Phys.Rev.C66:027902,2002.
    nucl-th/0112042;
Axel Drees, Haidong Feng, Jiangyong Jia,
Phys.Rev.C71:034909,2005
e-Print Archive: nucl-th/0310044
\bibitem{Pantuev:2005jt}
  V.~S.~Pantuev,
  %``Jet absorption and corona effect at RHIC: Extracting collision geometry
  %from experimental data,''
   hep-ph/0506095.
  %%CITATION = HEP-PH 0506095;%%


\bibitem{GM}D.~Molnar and M.~Gyulassy,
Nucl.\ Phys.\ {\bf A697} (2002)  495;
[Erratum-ibid.\ {\bf A703} (2002) 893]
%[ nucl-th/0104073].
%%CITATION = NUCL-TH 0104073;%%


\bibitem{Muller:2005en}
  B.~Muller and K.~Rajagopal,
  %``From entropy and jet quenching to deconfinement?,''
   hep-ph/0502174.
  %%CITATION = HEP-PH 0502174;%%

\end{thebibliography}
\end{document}